# Estimation of Channel Parameters in a Multipath Environment via Optimizing Highly Oscillatory Error–Functions Using a Genetic Algorithm

Amir Ebrahimi[#1], Ardavan Rahimian[*2]

[#]Department of Electrical Engineering, Iran University of Science and Technology, Tehran, Iran
[*]Electronic Engineering Group, International Branch, Amirkabir University of Technology, Tehran, Iran

[1]amir_ebrahimi@ee.iust.ac.ir
[2]rahimian@ieee.org

*Abstract*— **Channel estimation is of crucial importance for tomorrow's wireless mobile communication systems. This paper focuses on the solution of channel parameters estimation problem in a scenario involving multiple paths in the presence of additive white Gaussian noise. We assumed that number of paths in the multipath environment is known and the transmitted signal consists of attenuated and delayed replicas of a known transient signal. In order to determine the maximum likelihood estimates one has to solve a complicated optimization problem. Genetic Algorithms (GA) are well known for their robustness in solving complex optimization problems. A GA is considered to extract channel parameters to minimize the derived error–function. The solution is based on the maximum–likelihood estimation of the channel parameters. Simulation results also demonstrate GA's robustness to channel parameters estimation errors.**

## I. INTRODUCTION

The problem of channel estimation is of considerable interest in data communication and related fields [1] and it is a major challenge in space–time communication in a wireless environment [2]. Multipath parameter estimation is of interest in a number of communications applications. The phenomenon of specular multipath occurs when a single transmitted signal arrives at the receiver via any combination of several distinct, unequal length paths. The received signal is not just the transmitted signal, but several delayed and amplitude–weighted versions of the original transmitted waveform [3], the problem of an acoustic signal arriving at a receiver by more than one route and corrupting the original signal is common to most forms of SONAR. It is usually known as multipath and frequently involves reflections from sea bottom, volume, surface or some environmental phenomena [4]. Multipath reception of signals may be due to reflections at physical boundaries, refraction, scatterers, or layered transmission media [5].The problem of interest in this paper is characterized using the following model:

$$r(t) = \sum_1^M a_k s(t - \tau_k) + \widetilde{\omega}(t). \quad 0 \leq t \leq T \quad (1)$$

where:
$s(t)$     transmitted signal (pulse)
$a_k$     amplitude (attenuation value) for path $k$
$t_k$     time delay for path $k$
M     number of different paths
$\omega(t)$     white Gaussian noise.

In spite of existence of several works dealing with channel estimation, very few works analyses the channel estimation from the point of view of heuristic approaches [6]. In this paper, attempts are made to estimate the multipath parameters using a GA. The outline of this paper is as follows: Section II presents an introduction to Genetic–Algorithms. Section III develops the relationship between the "real–amplitude error–function" (RAEF) and "complex–amplitude error–function" (CAEF). Simulation results are presented in section IV. In section V we conclude this paper with some discussions.

## II. OVERVIEW OF GENETIC ALGORITHMS

Genetic Algorithms are non–determinist and chaotic search methods that use real world models to solve complex and at times intractable problems. In this method the optimal solution is found by searching through a population of different feasible solutions. After the population is studied in each iteration, the elicits are selected and are moved to the next generation under genetic operators. After adequate generations, better solutions dominate the search space therefore the population converges towards the optimal solution. Genetic algorithm functionality is based upon Darwin's theory of evolution through natural and sexual selection. As weaker elements are overthrown in nature, weaker solutions are omitted here. The best solutions of the last generation under the crossover operator usually direct the search space toward better answers. Mutation is considered as an important secondary genetic operator and effect the final solution.

In Genetics, attribute transfer in a gene sequence is done through common factors (Building Blocks). Consequently in genetic algorithms the parts of the genetic answer which have an effective role in the final solution are called the building block. The aim of genetic operator exertion on the population is to rearrange the building blocks placement to form effective compositions. Each element is usually built from a γ length binary string; therefore the search space will be $2^\gamma$. The initial population is randomly selected, although there may be different methods for specific problems but random selection is usually preferred to keep a suitable balanced distribution in the initial population. Each element is actually a codified version of the possible solution situated in a vast population. The genetic algorithm population evolves based on an evaluation function, selection, crossover, and mutation operators. The necessity to consider algorithm execution time and on the other hand reaching the global solution; makes correct population size selection an important task. Small population size may omit suitable solutions containing effective building blocks while big populations will increase the algorithm execution time and will waste computational resources. There should always be a trade off between population size and the two mentioned factors. Each element has a fitness value associated with it.

This factor shows the extent of influence which the element has on the final solution. As genetic algorithms usually act as optimizers, this factor can estimate the satisfaction percentage of the result. Using the fitness value in the selection operator, elements with higher value will be selected for crossover, mutation and some times to be transferred directly to the next generation. Crossover is the dominant operator in genetic algorithms. The typical model of this operator randomly selects two different elements from the population and substitutes these elements from one point in the string. For example two sequences like A1 and A2 can be considered as:
A1 = 0110 1111
A2 = 1110 0000
There are γ points for crossover in an element with the length of γ. In this example point number 4 has been selected. With the application of crossover the elements for the next generation will be as follows:
A′1 = 0110 0000
A′2 = 1110 1111

Different versions of this operator can be found in various texts. The given example is of a single point crossover although two point or even n point crossovers can be used. It's crucial to notice that with the increase in the number of points used in crossover, the effective building blocks may be divided and therefore ruined. Mutation is usually considered as a complementary operator. Its main purpose is to broaden the search space because using crossover alone may drive our results to one direction. For example having 11011 and selecting point 3 for mutation will result in 11111. As in nature, the probability of crossover application (Pc) is relevantly high where on the other hand mutation probability (Pm) is much lower. Many different methods have been proposed for initial genetic algorithm parameter setting. Dejong offered an initial population size of 50, 0.6 for Pc and 0.001 for Pm. Different approaches can be taken to terminate a genetic algorithm's execution. One of the ways is to end the execution after a fixed amount of iterations. Another approach is to end the algorithm when the best fitness value of the elements in the population has reached a specific peak, or in other cases the execution can be terminated when all the elements are the same however this is the case when mutation is not used [7].

**III. THE COMPLEX AND REAL AMPLITUDE ERROR FUNCTIONS**

In practice, we have only the samples of the received and transmitted signals for the time interval $0 < t < T$. The sample received waveform can be modelled as:

$$r(nT_s) = \sum_{k=1}^{M} a_k s(nT_s - \tau_k) + \omega[n], \quad (2)$$
$$0 \leq n \leq N-1$$

where the sampling interval $T_S = T/N$ and $W[n]$'s are the noise samples of the low–pass filtered version of the white noise $\tilde{\omega}(t)$. We assume that the autocorrelation of the white noise in the continuous time model is $R_{\tilde{\omega}}(\tau) = \sigma_{\tilde{\omega}}^2 \delta(\tau)$. Further, we assume that $\tilde{\omega}(t)$ is low–pass filtered by $h_{LP}(t) = (1/\pi t)\sin(\pi t/T_s)$ to yield $\omega(t)$ such that the discrete–time signal $\omega[n] = \omega(nT_s)$ has a flat power spectrum and that $\sigma_\omega^2 \equiv E\{\omega^2[n]\} = \sigma_{\tilde{\omega}}^2 / T_s$. The least–squares estimator of $a = [a_1, ..., a_M]^T$ and $\tau = [\tau_1 ... \tau_M]^T$ is obtained by minimizing the squared error function:

$$E_r(\tau,a) = \sum_{n=-N/2}^{N/2-1} \left| R[n] - S[n]\sum_{k=1}^{M} a_k e^{-j\tau_k 2\pi n/NT_s} \right|^2 \quad (3)$$

where $R[n]$ and $S[n]$ are the DFTs of $r[nT_s]$ and $s[nT_s]$, respectively. If the noise is white Gaussian, then the LS estimator is also the maximum–likelihood estimator (MLE). While minimizing the above error function, the conjugate symmetry structure of the spectra ($R[n]$, $S[n]$) implicitly constrains the optimal amplitudes to be real valued. Since the spectra in (3) are conjugate symmetric, let us consider a different error function that uses the spectra for only the positive frequencies.

$$E_c(\tau,a) = \sum_{n=0}^{N/2-1} \left| R[n] - S[n]\sum_{k=1}^{M} a_k e^{-j\tau_k 2\pi n/NT_s} \right|^2 \quad (4)$$

Unlike the previous case, the optimal $a$ that minimizes $E_c(a,\tau)$ can be shown to be complex valued. We refer to $E_r(a,\tau)$, which is the error function that constrains the amplitudes to be real valued, as the RAEF and $E_c(a,\tau)$, which is the error function that allows the amplitudes to be complex valued, as the CAEF. As a consequence of sampling, both of these error expressions have a periodicity of $T$ with respect to each $\tau_k$. In the above minimization, we threshold the signal spectrum and consider only those regions that are above a given value. This has the advantage of working with regions having good SNR's only. A rule of thumb is to set this threshold to roughly one twentieth to one tenth the peak magnitude of the signal spectrum. We choose the smoother error expression in (4) and modify it as (5) here:

$$E_c(\tau,a) = \sum_{n \in \mathbb{N}} \left| R[n] - S[n]\sum_{k=1}^{M} a_k e^{-j\tau_k 2\pi n/NT_s} \right|^2$$
$$= \|\tilde{r} - \tilde{p}(\lambda)a\|^2 \quad (5)$$

where
$\mathbb{N} = \{0 \le n \le N/2-1 : S[n] > \text{threshold}\}$
$= \{q_1 \ldots q_L\}$
$\lambda = [\lambda_1, \lambda_2 \ldots \lambda_M]^T$ with $\lambda_k = -\tau_k 2\pi/NT_s$
$r = [R[q_1] R[q_2] \ldots R[q_L]]^T$
$S = \text{diag}(S[q_1], S[q_2] \ldots S[q_L])$

$$A(\lambda) = \begin{pmatrix} e^{j\lambda_1 q_1} & e^{j\lambda_2 q_1} & \ldots & e^{j\lambda_M q_1} \\ e^{j\lambda_1 q_2} & e^{j\lambda_2 q_2} & \ldots & e^{j\lambda_M q_2} \\ \vdots & \vdots & \vdots & \vdots \\ e^{j\lambda_1 q_L} & e^{j\lambda_2 q_L} & \ldots & e^{j\lambda_M q_L} \end{pmatrix}$$

$$\tilde{p}(\lambda) = SA(\lambda) \quad (6)$$

$S[qk]$ and $R[qk]$ are those samples that lie above the threshold. The notation $n \in \mathbb{N}$ means that the summation is over those values of *n* that belong to the set N and not necessarily over contiguous values:

$$\tilde{p}(\lambda) = [p(\lambda_1) p(\lambda_2) \ldots p(\lambda_M)] \quad (7)$$

We have used a GA aimed at minimizing the error function in Equation (6) for the unknown time delays and attenuations.

## IV. SIMULATION EXAMPLE

The transmitted signal *s(t)* is a windowed linear FM signal. Its samples are given by:

$$S[n] = \begin{cases} a[n]\sin(2\pi(an^2+bn)) & n = 0 \ldots N-1 \\ 0 & \text{otherwise} \end{cases} \quad (8)$$

where
$$a[n] = \begin{cases} 0.5 - 0.5\cos(\pi n/N_w) & n = 0 \ldots N_w - 1 \\ 1.0 & N_w \ldots N - N_w - 1 \\ 0.5 - 0.5\cos(\pi(n-N)/N_w) & N - N_w \ldots (N-1) \end{cases} \quad (9)$$

is a window function, and $N=750$, $N_\omega = N/10$, $a=(f2-f1)/2N$, $b=f1, f1=0.1$, and $f2=0.15$. A synthetic three–path received signal is generated as:
$r[n] = s[n-200] - 0.8\, s[n-204] + 0.4\, s[n-220]$
$+ \omega[n]. \quad n = 0 \ldots 999. \quad (10)$

Now if we find the FFT of $r[n]$ and $s[n]$ and substitute them into (4), we obtain:
$E_c(a,\tau) = E_c(a_1, a_2, a_3, \tau_1, \tau_2, \tau_3) \quad (11)$

According to (10), the global minimum of $E_c(a_1, a_2, a_3, \tau_1, \tau_2, \tau_3)$ accures at the optimal values ($a_1=1, a_2=-0.8, a_3=0.4, \tau_1=200T_s, \tau_2=204T_s, \tau_3=220T_s$).

Fig.1 shows $E_c$ as a function of, $\tau_1$, when the values of the other parameters were set to their optimal values as follows:
($\tau_2=204T_s, \tau_3=220T_s, a_1=1, a_2=-0.8, a_3=0.4$)

Fig.2, 3 show $E_c$ as a function of, $\tau_2$ and $\tau_3$ respectively, when the values of the other parameters were set to their optimal values.

Fig.4, 5 and 6 show Ec as a function of, a1, a2 and a3 respectively, when the values of the other parameters were set to their optimal values;

From Fig.1, 2 and 3 it can be seen that $E_c$ is an oscillatory function, Hence, $E_c$ is chosen to be our fitness function for Genetic Algorithms, We can rewrite (11) as:
$E_c(a, \lambda) = E_c(a_1, a_2, a_3, \lambda_1, \lambda_2, \lambda_3) \quad (16)$

where,

$$\lambda_k = -\tau_k 2\pi / NT_s \qquad (17)$$

The global minimum of $E_c$ accures at the optimal values of (a1=1, a2=-0.8, a3 = 0.4, $\tau_1$=200 Ts, $\tau_2$=204Ts, $\tau_3$=220 $T_s$).

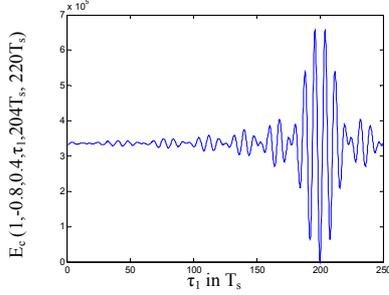

Fig. 1. $E_c$ an oscillatory function of the first time delay $\tau_1$

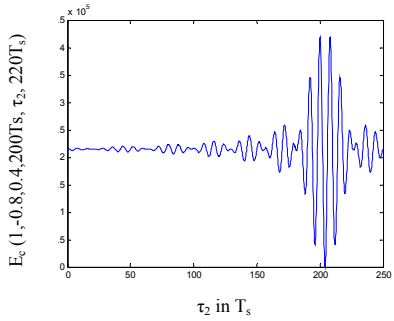

Fig. 2. $E_c$ an oscillatory function of the second time delay $\tau_2$

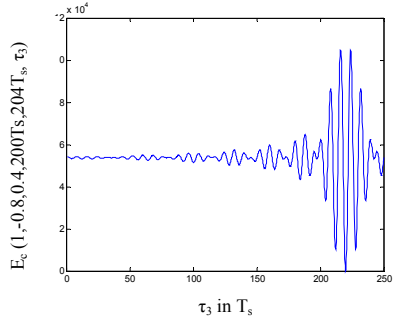

Fig. 3. $E_c$ an oscillatory function of the third time delay $\tau_3$

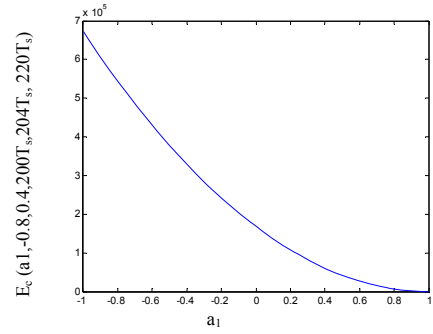

Fig. 4. $E_c$ a function of the first attenuation $a_1$

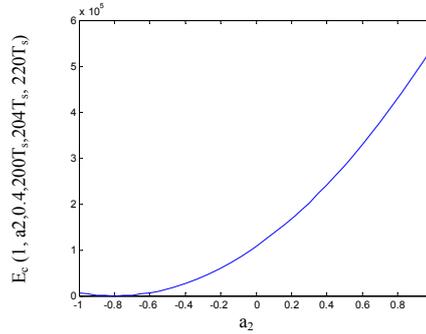

Fig. 5. $E_c$ a function of the second attenuation $a_2$

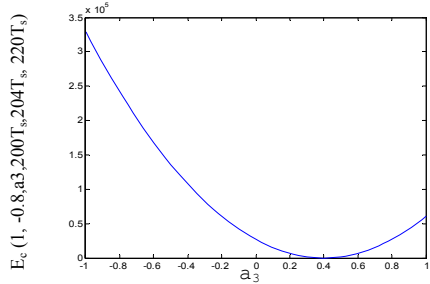

Fig. 6. $E_c$ a function of the third attenuation $a_3$

Fig. 4,5 and 6 shows Ec as a function of, a1, a2 and a3 respectively, when the values of the other parameters were set to their optimal values;

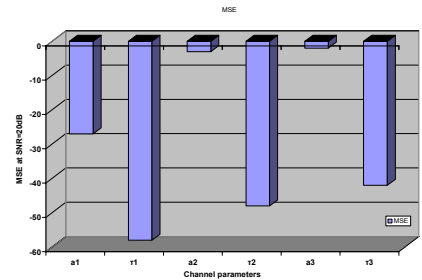

Fig. 7. MSE of the channel parameters at SNR=20dB

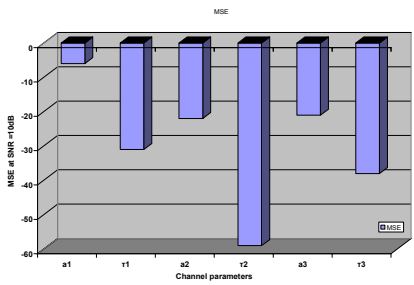

Fig. 8. MSE of the channel parameters at SNR=10dB

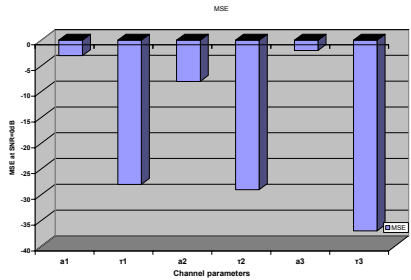

Fig.9. MSE of the channel parameters at SNR= 0dB

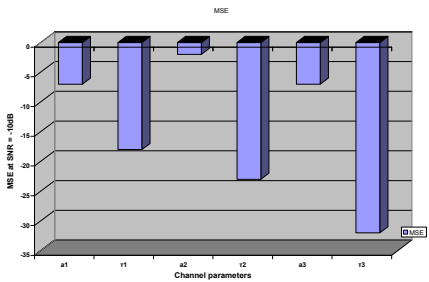

Fig. 10. MSE of the channel parameters at SNR= -10dB

The power of r[n] was fixed and the random noise sequences were scaled to obtain the required SNR. The figure of merit used in our simulations is the mean square error (MSE) in channel parameters estimation.

## V. CONCLUSION

We have investigated the application of the GA to the problem of channel parameters estimation in the presence of noise; the noise model is assumed additive white gaussian noise. We have used an error function which is a LS estimator. This error function is minimized for the unknown time delays and attenuations. The usefulness of the algorithm is demonstrated by an example. The algorithm's performance is considered under different SNR s. MSE's of the estimated channel parameters are derived. GA's performance is verified by the computer simulations. Many existing optimization algorithms have the following drawbacks:

i. They are likely to converge to some local minimum and they are very sensitive to the initial estimates of the parameters such as alternating projection (AP) and coordinate descent (CD) algorithms.

ii. Number of iterations is much more than GA such as AP and expectation maximization (EM).

The results show that GA can achieve satisfactory channel parameters estimation performance.